# An Impartial Perspective on Superconducting Nb$_3$Sn coated Cu RF Cavities for Future Accelerators

E. Barzi*

*Fermi National Accelerator Laboratory, Batavia, IL 60510, USA and Ohio State University, Columbus, OH 43210, USA*

B. C. Barish

*California Institute of Technology, Pasadena, CA 91125, USA and U.C. Riverside, Riverside, CA 92521, USA*

R. A. Rimmer

*Thomas Jefferson National Accelerator Facility, Newport News, VA 23606, USA*

A. Valente-Feliciano

*Thomas Jefferson National Accelerator Facility, Newport News, VA 23606, USA*

C. M. Rey

*Energy to Power Solutions (e2P), Tallahassee, FL 32304, USA*

W. A. Barletta

*Massachusetts Institute of Technology, Cambridge, MA 02139, USA*

E. Nanni

*SLAC National Accelerator Laboratory, Menlo Park, CA 94025, USA*

M. Nasr

*SLAC National Accelerator Laboratory, Menlo Park, CA 94025, USA*

M. Ross

*SLAC National Accelerator Laboratory, Menlo Park, CA 94025, USA*

M. Schneider

*SLAC National Accelerator Laboratory, Menlo Park, CA 94025, USA*

S. G. Tantawi

*SLAC National Accelerator Laboratory, Menlo Park, CA 94025, USA*

P. B. Welander

*SLAC National Accelerator Laboratory, Menlo Park, CA 94025, USA*

E. I. Simakov

*Los Alamos Accelerator Laboratory, Los Alamos, NM 87545, USA*

I. O. Usov

*Los Alamos Accelerator Laboratory, Los Alamos, NM 87545, USA*

L. Alff

*Technische Universität Darmstadt, Darmstadt 64289, Germany*

N. Karabas

*Technische Universität Darmstadt, Darmstadt 64289, Germany*

M. Major

*Technische Universität Darmstadt, Darmstadt 64289, Germany*

J. P. Palakkal

*Technische Universität Darmstadt, Darmstadt 64289, Germany*

S. Petzold

*Technische Universität Darmstadt, Darmstadt 64289, Germany*

*barzi@fnal.gov – DISCLAIMER: The authors are indeed striving to be impartial. With this paper, for instance, the SLAC team with Cu tradition is clearly sending the message that they are not partial to Cu.

N. Pietralla
*Technische Universität Darmstadt, Darmstadt 64289, Germany*
N. Schäfer
*Technische Universität Darmstadt, Darmstadt 64289, Germany*
A. Kikuchi
*National Institute for Materials Science, Tsukuba, Ibaraki 305-0047, Japan*
H. Hayano
*High Energy Accelerator Research Organization (KEK), Tsukuba, Ibaraki 305-0801, Japan*
H. Ito
*High Energy Accelerator Research Organization (KEK), Tsukuba, Ibaraki 305-0801, Japan*
K. Umemori
*High Energy Accelerator Research Organization (KEK), Tsukuba, Ibaraki 305-0801, Japan*
H. Monjushiro
*High Energy Accelerator Research Organization (KEK), Tsukuba, Ibaraki 305-0801, Japan*
T. Kubo
*High Energy Accelerator Research Organization (KEK), Tsukuba, Ibaraki 305-0801, Japan*
H. Hama
*Tohoku University, Sendai, Miyagi 980-8577, Japan*
S. Kashiwagi
*Tohoku University, Sendai, Miyagi 980-8577, Japan*
F. Honda
*Kyushu University, Nishiku, Fukuoka 819-0395, Japan*
K. Takahashi
*Akita Kagaku Co., Ltd., Nikaho, Akita 018-0402, Japan*
R. Takahashi
*Akita Kagaku Co., Ltd., Nikaho, Akita 018-0402, Japan*
Y. Kondo
*Akita Industrial Technology Center, Araya, Akita 010-1623, Japan*
K. Yamakawa
*Akita Industrial Technology Center, Araya, Akita 010-1623, Japan*
K. Kon
*Iwate Industry Promotion Center, Kitaiioka, Morioka 020-0857, Japan*

ABSTRACT

This Snowmass21 Contributed Paper encourages the Particle Physics community in fostering R&D in Superconducting $Nb_3Sn$ coated Copper RF Cavities instead of costly bulk Niobium. It describes the pressing need to devote effort in this direction, which would deliver higher gradient and higher temperature of operation and reduce the overall capital and operational costs of any future collider. It is unlikely that an ILC will be built in the next ten years with Nb as one of the main cost drivers of SRFs. This paper provides strong arguments on the benefits of using this time for R&D on producing $Nb_3Sn$ on inexpensive and thermally efficient metals such as Cu or bronze, while pursuing in parallel the novel U.S. concept of parallel-feed RF accelerator structures. A technology that synergistically uses both of these advanced tools would make an ILC or equivalent machines more affordable and more likely to be built. Such a successful enterprise would readily apply to other HEP accelerators, for instance a Muon Collider, and to accelerators beyond HEP. We present and assess current efforts in the U.S. on the novel concept of parallel-feed RF accelerator structures, and in the U.S. and abroad in producing $Nb_3Sn$ films on either Cu or bronze despite minimal funding.

# 1. EXECUTIVE SUMMARY AND RECOMMENDATIONS

## 1.1. Summary

The concept that particle physics is an international endeavor cannot be better conveyed than with the following. *This is driven in part by the universal nature of the scientific goals and achievements, but also by the growing experimental challenges of the field. No one institution or nation can assemble the resources or expertise needed to explore the frontiers of the field without help from the international community. Nor would it want to. Scientists in particle physics understand that the global nature of the field is one of its greatest strengths. The diversity of national, social and cultural backgrounds present in the experiments and labs enriches the pool of intellectual thought and solidifies the validity of their scientific findings.* [1]

This fact applied to the story of the ILC [2] is one of the primary messages of this paper. No one institution or nation could yet assemble the 5.5 billion ILCU (an ILC unit is 1$ at 2012 value) capital cost and the 0.3 billion ILCU yearly operational costs of a 250 GeV electron-positron collider. And this is for a machine that, perhaps as never before in HEP history, was so unanimously supported. Since the 1990s, ICFA as well as numerous national HEP communities and government officials, advocated for it, and this past May 2021, the CERN Council approved and published the European strategy which, once again, put an electron-positron Higgs factory with c.m. energy of 250 GeV as the highest-priority next collider. A large fraction of the capital cost is in the superconducting and heavily processed bulk Nb SRF cavities, and in the cryogenic plant required to cool these cavities down to 2 K superfluid Helium, which also impacts the operational cost.

Since the decision in 2003 of the International Technology Recommendation Panel (ITRP) to focus on SRF, the technology in Normal-Conducting RF (NCRF) has made considerable progress [3], achieving linac gradients exceeding 160 MV/m with cryo-cooled Cu. $C^3$ is a concept that is aimed at developing NCRF accelerator technology to operate at high gradient with high RF-to-beam efficiency. The two principal innovations for the $C^3$ concept are: (1) the use of highly-optimized reentrant cells with distributed coupling to power the linac without cell-to-cell RF coupling, and (2) the operation of the Cu accelerating structure at liquid Nitrogen temperatures (77~K). The necessary structure is machined in two halves by low-cost numerically-controlled milling machines. This machining process produces ultra-high vacuum (UHV) quality surfaces that need no further machining before a standard Cu surface etch. This manufacturing technique provides an ideal Cu surface to be coated with superconducting films, as it allows complete access to the inner cavity surface for the coating process. The system is then assembled simply by joining the two blocks.

SRF cavities with a thin layer of $Nb_3Sn$ coated onto their inner surface should produce accelerating gradients on the order of 100 MV/m, twice that expected for Nb cavities. With a $T_{c0}$ double that of Nb, they also produce a cavity quality factor $Q_0$ about 30 times larger than for Nb cavities. The larger $T_{c0}$ of $Nb_3Sn$ has also the advantage of allowing the cavities to operate at 4.5 K rather than in superfluid helium at 2 K that is used for bulk Nb cavities to obtain a higher gradient. Unfortunately, most of the funding on $Nb_3Sn$ in the U.S. has been on a Sn vapor diffusion process on the internal surface of a Nb bulk cavity proposed more than 40 years ago [4]-[8]. At the time, 1.5 GHz single and multi-cell Nb cavities coated with $Nb_3Sn$ were investigated up to peak accelerating fields of ~15-30 MV/m. Progress in this process in its last revival decade has plateaued to ~20-24 MV/m due to various limitations [9]. Moreover, it cannot be used on Cu as it requires a temperature of ~1100°C, which is higher than Cu melting point. On the other hand, there are small R&D efforts in the U.S. and abroad in producing Nb$_3$Sn films on either Cu or bronze. These efforts should be encouraged and sponsored at the global level, and at the very least within the U.S.. A machine whose RF structures are made of bulk Cu or bronze that operates with the properties of Nb$_3$Sn superconductor would deliver higher gradient and higher temperature of operation and reduce the overall capital and operational costs of any future collider.

## 1.2. Recommendations

Based on the assessment of current efforts in the U.S. on the novel concept of parallel-feed RF accelerator structures, and in the U.S. and abroad in producing Nb$_3$Sn films on either Cu or bronze, we recommend that the Particle Physics community foster R&D in Superconducting Nb$_3$Sn coated Cu RF Cavities instead of costly bulk Nb. CERN and other European and Japanese institutions had started pursuing Nb$_3$Sn for SRF applications, with the ultimate goal of finding a coating technique to use on Cu cavities, before the U.S.. It is unlikely that an ILC will be built in the next ten years with Nb as one of the main cost drivers of SRFs. A devoted global effort in developing Cu cavity structures coated with Nb$_3$Sn would make the ILC or Higgs factories more affordable and more likely to be built. Using the next decade for R&D on producing Nb$_3$Sn on inexpensive and thermally efficient metals such as Cu or bronze, while pursuing in parallel the novel U.S. concept of parallel-feed RF accelerator structures, would compound the best of both worlds. Not only do parallel-feed RF structures enable both higher accelerating gradients and higher efficiencies, but they would be applicable to both Cu and Nb$_3$Sn coated Cu cells. Increased effort on these two techniques would synergize expenditures towards 10-year progress, which will naturally converge to a clear decision by the community on which path to take for the RF of an ILC or any future accelerator. If for any reason, the C$^3$ structures were not ready in ten years, the current methods of Nb$_3$Sn coatings on Cu or bronze are geared towards standard cavity cells. Were one to succeed, it could still be implemented on conventional Cu RF. In conclusion, the use of distributed coupling structure topology within improved performance parameters together with Nb$_3$Sn coating technology can lead to a paradigm shift for superconducting linacs, with higher gradient, higher temperature of operation, and reduced overall costs for any future collider.

## 2. INTRODUCTION

Despite the unanimous global consensus, completion of the TDR for a 500 GeV collider by 2013, and reduction of the energy and the cost by two in 2017, the 5.5 billion ILCU capital cost and the 0.3 billion ILCU yearly operational costs have been all along considered problematic. This is now further exacerbated by concerns on the economy due to Covid and the war in Ukraine. The production chain for ILC's Nb SRF cavities includes a long series of processes, such as electron beam welding, high-pressure rinsing, electro-polishing, furnace heat treatments. Although projects like CEBAF, SNS, XFEL, LCLS-II and ESS have produced scale-up capabilities, cost-driving requirements remain in the materials specifications and fabrication. The additional complexity of surface doping/infusion with nitrogen and its associated requirements have increased process costs of Nb SRF cavities. It is not clear whether the saving in operation costs will offset such a capital increase in a reasonable time.

Since the decision in 2003 of the International Technology Recommendation Panel (ITRP) to focus on SRF, the technology in Normal-Conducting RF (NCRF) has made considerable progress [3], achieving linac gradients exceeding

160 MV/m with cryo-cooled Cu. $C^3$ is a concept that is aimed at developing NCRF accelerator technology to operate at high gradient with high RF-to-beam efficiency.

SRF accelerators must therefore also make progress to remain competitive. Prior to and within the European EuCARD-2 funded R&D project for next-generation accelerators, CERN and other European and Japanese institutions started pursuing $Nb_3Sn$ for SRF applications with the ultimate goal of finding a coating technique to use on Cu cavities. The lower cost and thermal efficiency of the Cu are key reasons for CERN, for instance, to choose Cu cavities lined with Nb for its accelerators. The two most important properties for SRF performance are the accelerating field $E_{acc}$ and the cavity quality factor $Q_0$. The accelerating gradient in SRF cavities is proportional to the peak magnetic field on the cavity wall. At RF frequencies, $E_{acc}$ is limited by the peak magnetic field reaching or exceeding the superheated critical magnetic field, $H_{sh}$. The maximum accelerating gradient expected for Nb cavities is ~50 MV/m. With a theoretical $H_{sh}$ of 0.42 T, as compared to 0.25 T for Nb, SRF cavities with a thin layer of $Nb_3Sn$ coated onto their inner surface should produce accelerating gradients on the order of 100 MV/m for a TESLA shape cavity [10]-[12]. With a higher $T_{c0}$ of up to 18 K, as compared to a $T_{c0}$ of 9.2 K for Nb, SRF cavities coated with $Nb_3Sn$ also produce a cavity quality factor $Q_0$ about 30 times larger than for Nb cavities. The larger $T_{c0}$ of $Nb_3Sn$ has also the advantage of allowing the cavities to operate at 4.5 K rather than in superfluid helium at 2 K that is used for bulk Nb cavities to obtain a higher gradient. This would decrease capital costs for the cryogenic plant, as well as operation costs, by a substantial amount.

The necessary structure for the $C^3$ concept is machined in two halves by low-cost numerically-controlled milling machines. This machining process produces ultra-high vacuum (UHV) quality surfaces that need no further machining before a standard Cu surface etch. This manufacturing technique provides an ideal Cu surface to be coated with superconducting films, as it allows complete access to the inner cavity surface for the coating process. The system is then assembled simply by joining the two blocks. We present and assess current efforts in the U.S. on the novel concept of parallel-feed RF accelerator structures. This will be followed by examples of efforts by research groups in the U.S. and abroad on both thin and thick $Nb_3Sn$ films to be produced on Cu and/or bronze. It has to be noted that these coating methods are all scalable also to standard cavity cells, albeit with some additional effort.

## 3. ADVANCED NCRF LINAC CONCEPT FOR A HIGH ENERGY E$^+$E$^-$ LINEAR COLLIDER

### 3.1. State-of-the-art Normal-Conducting Cu RF Cavities

$C^3$ is a concept that is aimed at developing normal conducting RF (NCRF) accelerator technology to operate at high gradient with high RF-to-beam efficiency [3]. $C^3$ accelerators are bringing recent advances in the understanding of high-gradient operation [13], cavity design and RF power distribution [14], RF pulse compression [15], and cryogenic operation [16] to improve the performance of NCRF accelerators for high-gradient, high-brightness and high-luminosity applications. The two principal innovations for the $C^3$ concept are: (1) the use of highly-optimized reentrant cells with distributed coupling to power the linac without cell-to-cell RF coupling, and (2) the operation of the Cu accelerating structure at liquid Nitrogen temperatures (77~K) to increase the RF efficiency of the structure by a factor of three, while also increasing the strength of the material. This has been found to correlate with the achievable operating gradient. The $C^3$ approach has the potential of operating at extremely high gradients. Prototype structures have been operated with beam up to 160 MeV/m [16] and single cell test cavities have exceeded 200 MeV/m gradients [17]. The $C^3$ cryomodule that is under development for the main linac and the RF sources that accompany it fits comfortably within the existing diameter tunnel that is planned for the ILC main linac tunnel. The nominal operating parameters for $C^3$ technology are 120 MeV/m gradient with ~90% fill factor for the cryomodule that forms the basis of the main linac design. A higher gradient version at 155 MeV/m is also being explored. The expectation over the next decade is to push the accelerating gradient of the $C^3$ linac well beyond 120 MeV/m, measure the break down rate in realistic operating conditions and make a determination of the physical limit of the installed cryomodules capacity for higher gradient.

### 3.2. Parallel-feed RF Accelerator Structures

The implementation of new approaches to designing, building and operating a normal conducting accelerator structure has been key to the development of the technology for $C^3$. In the initial study of cavity shapes to maximize the on-axis accelerating field while minimizing breakdown, suitable shapes were found, but the iris was too small to propagate the fundamental cavity mode. This was overcome with distributed coupling of RF power, implemented with parallel manifolds feeding each cavity with the proper RF phase and fraction of the inlet RF power. It was realized that this relatively complex structure was possible to be machined in two halves (or 4 quarters) by low-cost numerically-controlled milling machines. This machining process produces ultra-high vacuum (UHV) quality surfaces that need no further machining before a standard Cu surface etch. Finally, the operating temperature of the Cu accelerator was reduced to both increase the electrical conductivity of the material and increase the material strength. In liquid Nitrogen at ~80 K, the increased conductivity of Cu at this temperature and the cavity shape optimization results in a shunt impedance of 300 M$\Omega$/m, i.e. 6 times the effective shunt impedance of the NLC. The increase in electrical conductivity also reduces the required RF power, the source of which is a costly and complex part of the linac infrastructure, and reduces the thermal stress in the material. Thermal stress in the copper surface during the RF pulse results in cyclic fatigue of the material, crystal growth, motion of dislocations in the material and finally in electrical breakdown.

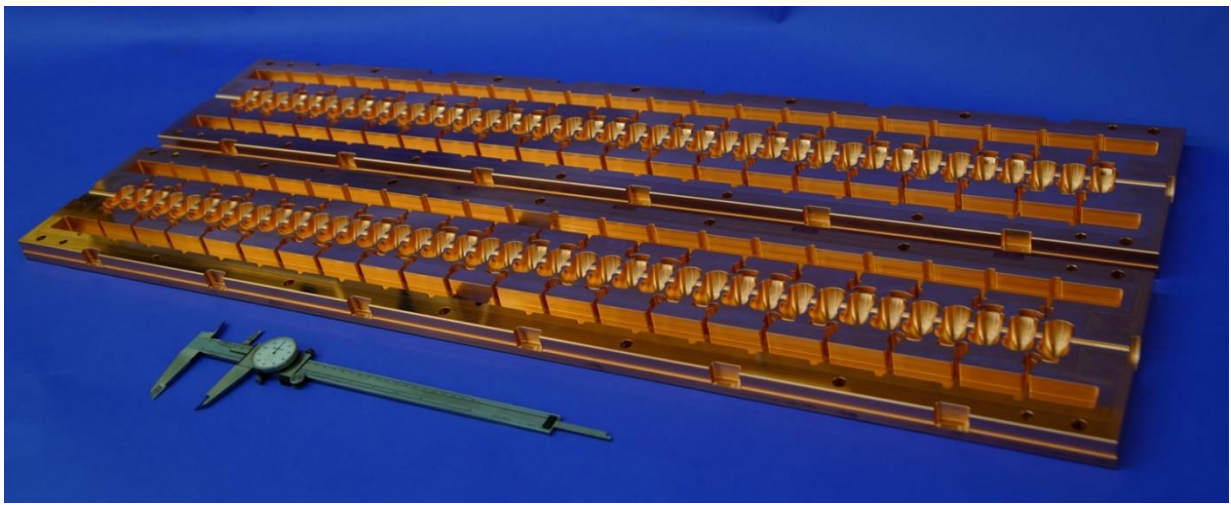

Figure 1: Meter-scale prototype $C^3$ structure.

The $C^3$ accelerating structure utilizes a Cu standing wave distributed coupling RF structure. The first meter-scale prototype $C^3$ structure is shown in Fig. 1. The aperture of the cavity is determined by considering short-range and long-range wakefield effects for the nominal bunch charge of 1 nC. The baseline phase advance between cells is $\pi$. The frequency of operation for the main linac is 5.712 GHz (C-band) in order to provide a high shunt impedance. Optimization of RF phase advance to $3\pi/8$ phase advance per cell, reduced cell length and proper power and phase RF manifold design for distributed coupling.

### 3.3 Application of Nb$_3$Sn Coated Cu Cavities in Parallel Feed Accelerator Structure

For high-β electron-positron accelerators where β =v/c ≈ 1, the best-known and most widely used SRF structure is the TESLA cavity [18]. The 1.3 GHz TESLA cavity and variations of it are the basis for the ILC, the XFEL, and the LCLS-II. The TESLA cavity is about 1.3 m long and consists of nine coupled elliptical cells. Power is fed to the cavity from one end by a coaxial input coupler, with the coupling strength between cells optimized in order to obtain the necessary dispersion and uniform field distribution.

Within the novel topologies developed at SLAC, genetic optimization algorithms that focus on efficiency or high gradient performance produce cavity shapes that are generally incompatible with either traveling-wave or standing-wave structures with a pre-defined coupling between cavities [19]. Instead of employing coupled cells, the new SLAC topology feeds each cell in the accelerator structure independently (see Section 3.2). Such a structure has been demonstrated with normal-conducting bulk Cu. The first superconducting prototype of such a structure is being tested at SLAC. This initial prototype is made of bulk Nb.

Table 1 shows a comparison between the TESLA Linac and a linac designed using this new distributed coupling topology. The cell shape for the distributed coupling topology was parameterized using a series of elliptical and straight segments as described in [20], and optimized for either maximum shunt impedance, $R_{sh}$, while constraining the maximum accelerating gradient, $E_{acc}$, relative to the peak surface magnetic field, $H_{pk}$. In all cases, the electric field ratio, $E_{pk}/E_{acc}$, was limited to a maximum value of about 2.0 (similar to the TESLA cavity). Table 1 lists several key parameters for this cell design and draws a direct comparison with the state-of-the-art TESLA cavity. Assuming both are fabricated from bulk Nb and operated at a temperature of 2 K, the optimized cell shape achieves both 60% less RF power dissipation and 70% greater accelerating gradient.

Table 1: Comparison between the SCPF and TESLA Cavities, assuming both are fabricated from Bulk Nb. At 2 K, $R_s$ is set to 27 nΩ (BCS).

| Parameter | SC-PF | TESLA |
|---|---|---|
| $f$ (GHz) | 1.3 | 1.3 |
| Mode (phase advance) | $2\pi/3$ | $\pi$ |
| $a$ (cm) (aperture radius) | 0.66 | 3.5 |
| $Q_0$ | 8.0e9 | 1.0e10 |
| $R_{sh}$ (Ω/m) | 2.4e13 | 9.8e12 |
| $E_{pk}/E_{acc}$ | 2.05 | 1.98 |
| $B_{pk}/E_{acc}$ (mT·m/MV) | 2.41 | 4.17 |
| $R_{sh}/Q_0$ (Ω/m) | 3050 | 983 |
| $G$ (Ω) | 217 | 271 |
| $G \cdot R_{sh}/Q_0$ (Ω$^2$) | 5.08e4 | 3.07e4 |
| $P_{loss}/E_{acc}^2$ (mW·m/MV2) | 41.1 | 102 |
| $U/E_{acc}^2$ (mJ·m/MV$^2$) | 40.2 | 124 |

Indeed, this new topology does not lend itself to the usual mechanical construction techniques for normal or superconducting structures. Hence, researchers at SLAC had to invent a new methodology for the manufacturing of such a structure. The so-called split cell manufacturing technique was developed to allow the construction of these cells from two blocks of metal [14]. Both the manifolds and the cavities have no currents crossing the plane, which splits the structure in half along the long dimension of the manifold cross-section. Manufacturing from two blocks reduces the complexity of the structure and provides logical positions for both the cooling tubes and the tuning holes. It also provides a possibility for low-temperature manufacturing techniques, including electron beam welding, and thereby allowing the use of unannealed Cu alloys and hence greater tolerance to high gradient operation. The circuit halves are aligned with an elastic averaging technique.

This manufacturing technique provides an ideal Cu surface to be coated with superconducting $Nb_3Sn$ films, as it allows complete access to the inner cavity surface for the coating process. Then, the system is assembled simply by joining the two blocks, which can be achieved using various techniques, including diffusion bonding.

In conclusion, the use of distributed coupling structure topology within improved performance parameters together with $Nb_3Sn$ coating technology can lead to a paradigm shift for superconducting linacs. It will allow the use of Cu-coated Superconducting $Nb_3Sn$ linac to achieve parameters listed in Table 1 but with operation at higher temperatures (~ 4K).

## 4. ELECTROPLATING OF $NB_3SN$

In wires, the superconducting $Nb_3Sn$ with A15 crystalline structure is formed through solid diffusion of the Sn into the Nb through the Cu matrix of the composite Nb/Sn/Cu wire. In the presence of Cu as ternary element, the maximum temperature of the heat treatment cycle needed for $Nb_3Sn$ formation is easily less than 700C [21]. When Cu is not present in the system, as for instance in the Sn vapor diffusion process, the temperature required from the binary phase diagram is much higher.

The most prized electroplating techniques, or any other method to form $Nb_3Sn$ coatings, are those with direct deposition of the $Nb_3Sn$ superconducting phase, where no heat treatment is required. One such method [22] was developed a few years ago within an Italian student program [23]. This electroplating technique allowed to directly deposit stoichiometric $Nb_3Sn$ on a Cu substrate from ionic solutions. However, due to lack of funding, reproducibility was never proven.

A simpler electroplating technique to coat Nb surfaces with Cu and Sn layers from aqueous solutions and produce $Nb_3Sn$ during a standard heat treatment was developed and made reproducible at FNAL in the last few years [24]-[26]. The know-how was then transferred to KEK within an U.S.-Japan Science and Technology Cooperation Program in HEP. The method was adapted at KEK on the basis of the chemicals commercially available in Japan. The electroplating technique herein described can be implemented on Cu surfaces too after sputtering them with Nb by using a simple magnetron system (see Section 5). The advantages of electro-deposition are its simplicity, accurate control, and low cost. It is an ideal technique to coat the large surfaces of Fig. 1.

In electro-chemical deposition the metallic coating is deposited on another metal surface through an electrolyte solution. The metal to be plated acts as the cathode and when the appropriate current is applied, positively charged ions traveling from the anode into the solution will discharge and get deposited on the cathode until a film of desired thickness is formed. Fig. 2 shows the sequence of deposited layers. First, using the Nb as cathode and Cu anode, a thin seed layer of Cu is deposited in an acid solution. The Cu lowers the formation temperature for the A15 compound and suppresses the unwanted NbSn$_2$ and Nb$_6$Sn$_5$ phases. In a second electroplating step, the resulting Nb/Cu sample is used as cathode, and a thick layer of Sn is deposited with a Sn anode within a commercial Sn-rich solution. And finally, on the resulting Nb/Cu/Sn sample, a Cu layer is again deposited using a Cu anode. Each electrodeposition step is carried out at near room temperature and at atmospheric pressure. Details are provided in [24].

Several superconducting Nb$_3$Sn films of thickness of up to 12 μm were obtained on Nb substrates of 0.3 mm to 0.5 mm thickness by studying and optimizing most parameters of the electro-plating process, including:

- Bath composition and anode materials for each of the three deposition steps;
- Current densities, deposition times, stirring rates, and cathode and anode relative orientation in DC mode;
- Current densities, deposition times, stirring rates, cathode and anode relative orientation, pulse frequencies, and duty cycles in pulsed mode.

Nb$_3$Sn is then formed through solid diffusion by heat treating the multi-layered samples in inert atmosphere (argon) at a maximum temperature of 700°C. After reaction, the Cu and bronze phases formed on the outer surface of the Nb$_3$Sn are removed with aqua regia.

Fig. 3 (a) shows the $T_{c0}$ result of 17.5 K obtained for a sample produced and measured resistively at FNAL. The same sample was also measured by SQUID at NIMS and the resulting $T_{c0}$ was 17.6 K. The two results are remarkably consistent. Fig. 4 shows the *M-H* curve obtained for this sample with the SQUID magnetometer. The lower critical field $H_{c1}$(4.2K) for this sample was 600 Oe. Fig. 5 shows the DC test at KEK of $T_{c0}$ = 17.5 K of a different Nb$_3$Sn film sample produced at KEK.

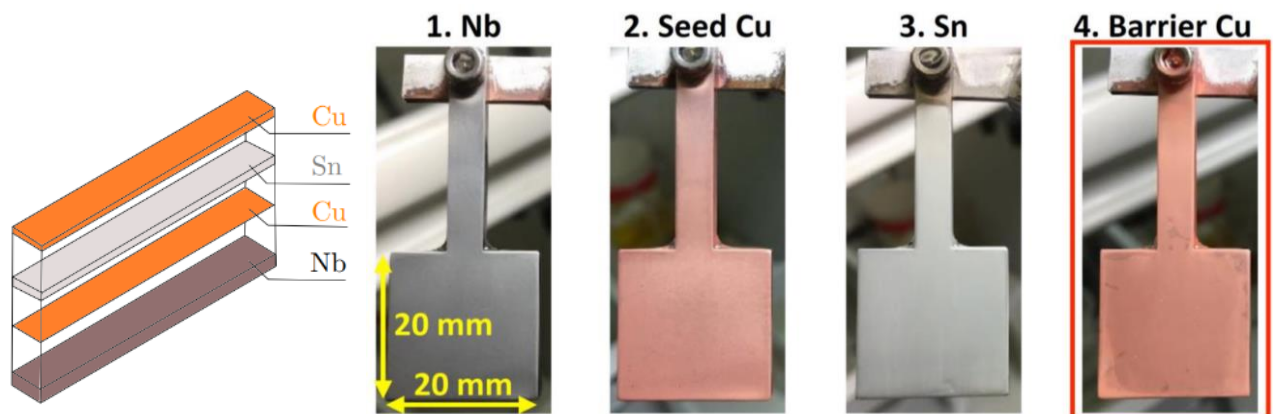

Figure 2: Sequence of deposited layers (left), and pictures of sample at each deposition step (right).

For the next stage, electroplating research in Japan was expanded with the collaboration with Akita Kagaku Co. Ltd, an electroplating company. The collaboration also expanded to Tohoku University, Akita Industrial Technology Center. Re-optimization of pre-treatment chemicals, plating chemicals, Nb purity (99.83% Nb and RRR=300 Nb), and annealing temperature brought the $T_{c0}$ for 99.83% Nb up to 17.8K, compared with 17.5K for a Nb of 300 in RRR. A 3 GHz Nb small cavity was prepared for Nb$_3$Sn electroplating inside. The RF performance ($Q_0$ and $E_{acc}$) of the 3 GHz Nb cavity was tested before plating at liquid He temperature (4.2K).

The electroplating technique herein described can be implemented on Cu surfaces too after sputtering them with Nb by using a simple magnetron system (see Section 5). Fig. 6 shows the results obtained at NIMS of sputtering 1 μm of Nb layer (right) on 3 mm thick Oxygen-free Cu plates that are 17 mm wide (left).

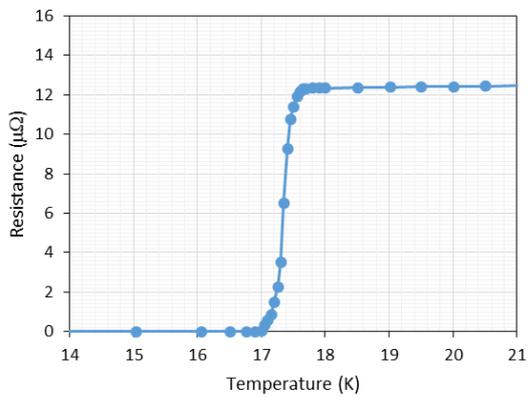

Figure 3: (a) DC test at FNAL of $T_{c0} = 17.5$ K of a Nb$_3$Sn film sample made at FNAL.

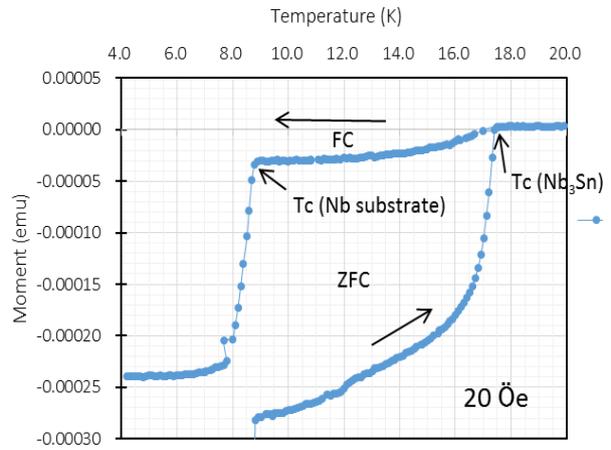

Figure 3: (b) *M-T* curve of same sample from Fig. 3 (a) obtained with SQUID magnetometer at NIMS. The $T_{c0} = 17.6$ K.

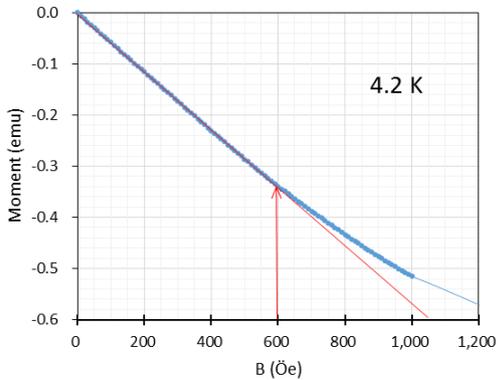

Figure 4: *M-H* curve of same sample from Fig. 3 (a) obtained with SQUID magnetometer at NIMS. The $H_{c1}(4.2\ K) = 600$ Oe.

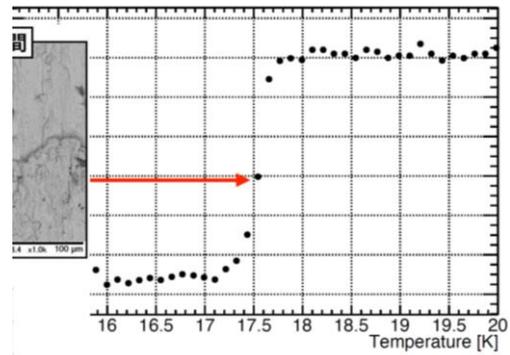

Figure 5: DC test at KEK of $T_{c0} = 17.5$ K of a Nb$_3$Sn film sample produced at KEK.

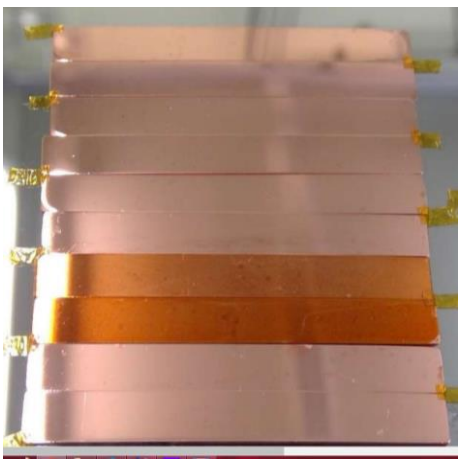
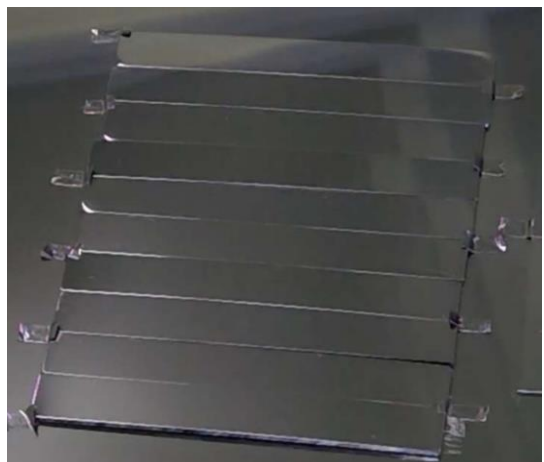

Figure 6: Oxygen-free Cu plates 3 mm thick and 17 mm wide before Nb sputtering (left) and after sputtering of a Nb layer 1 μm thick (right).

## 5. MAGNETRON SPUTTERING OF NB$_3$SN

Among the most promising approaches for Nb$_3$Sn film deposition on Cu is magnetron sputtering, which can be performed either (1) sequentially to form a multi-layer structure of Nb and Sn followed by post-reaction [27]; (2) in a co-sputtering [28] mode from two targets; (3) or from a single stoichiometric target [29], [30]. Using two separate targets in a co-sputtering setup allows tuning the kinetic energies of both elements independently. This process leads to the superconducting phase formation at much lower substrate temperatures as compared to thermal diffusion conditions. For instance, direct Nb$_3$Sn deposition was achieved on fused silica substrates by magnetron co-sputtering at 435°C, with a $T_{c0}$ of 16.3 K [28]. This approach paves the way for the use of Nb$_3$Sn as a coating in cryogenically efficient copper-based cavities avoiding the detrimental interdiffusion of Cu.

### 5.1. Magnetron Sputtering at the Technische Universität Darmstadt (TU Darmstadt)

This group at the Technische Universität Darmstadt has achieved low temperature co-sputtering of Nb$_3$Sn films on fused silica substrate. The Nb$_3$Sn superconducting phase was directly formed on the substrate at 435°C, without the need of further high temperature annealing. Detailed structural investigations were done. Phase pure films (XRD) of high homogeneity and good performance (*R-T* and *M-T* measurements) with critical temperature $T_{c0}$ = 16.3 K [28] were obtained.

Ongoing projects include the following:

**1. Further characterization of the Nb$_3$Sn by low-temperature synthesis**

Investigating the effect on sputtering power of low temperature growth (320°C) includes study of the crystal structure (via grazing angle XRD), morphology (SEM), role of grain boundaries. Resistivity vs. temperature measurements and investigating the magnetic field dependence of the current-voltage characteristics clarifies the role of grain boundaries on superconductivity. The film local chemical homogeneity and crystal quality are checked by Extended X-ray absorption fine structure (EXAFS) spectroscopy and micro- X-ray absorption (XAS). The results obtained so far indicate that a pure phase of textured Nb$_3$Sn films can be grown on fused silica. Higher sputtering power leads to sharper superconducting transition (with higher $T_{c0}$). This behavior is supported by the result of the current-voltage (*I-V*) measurements in magnetic field. The films with high sputtering power show no magnetic field dependence, linked to the homogenous grain-boundaries. In contrast the films sputtered at lower powers show field dependent *I-V* curves, which are explained with the grain-boundaries acting as Josephson-junctions (weak superconducting links), as first predicted in [31]. XAS and EXAFS measurements confirmed the homogeneity of the samples sputtered with high power (locally and globally) in contrast to the segregations found at low-power sputtered samples.

**2. Transfer of the Nb$_3$Sn sputtering process to Cu substrate**

Cu has a relatively high diffusivity. To successfully coat Cu cavities, the ideal temperature should be limited to a maximum of ~500 °C. This limitation is given by the interdiffusion of Cu in the Nb$_3$Sn thin film. For instance, ~0.6 at% Cu homogeneously distributed in the film cross section can be found at 520°C. Perhaps this Cu content is tolerable. However, ideally the process temperature should not be above this value. A critical temperature of 16.8 K (point of zero resistance) was reached in a phase pure film (as seen by XRD) at a process temperature of only 480°C. It is further shown that a stoichiometry slightly higher than 25 at% tin is not affecting shielding properties (shown by *M-T* measurements). The film roughness (as seen by profilometer, SEM and AFM) is higher than on fused silica as the starting Cu roughness is higher. Still, the roughness is about 100 nm which is low in comparison to other processes. Mechanical properties look very promising: scratch tests and nano-indentation tests showed excellent adhesion of the $Nb_3Sn$ films on Cu. The next step is the upscaling of the sputtering process to the cavity geometry.

## 5.2. Magnetron Sputtering at Los Alamos National Laboratory

At LANL an ion beam sputtering technique which offers superior control over the uniformity of a deposited film on internal surfaces of complex parts was developed. This coating technique was demonstrated on small coupons, and produced coupons coated with $Nb_3Ge$ with the correct stoichiometry. LANL also demonstrated coupons coated with $Nb_3Sn$ that showed superconducting transition at temperatures as high as 16K. The $Nb_3Sn$ coupons were produced at significantly lower temperatures compared to coatings manufactured at other National Laboratories (700°C vs. 1200°C).

Magnetron sputtering of $Nb_3Ge$ was performed on small 5x5 $mm^2$ coupons made of sapphire, Niobium, and carbon substrates. The first set of thin films was found to be Nb-depleted. All subsequently produced films had correct stoichiometry (75 at% of Nb and 25 at% of Ge). X-ray diffraction (XRD) and Transmission Electron Microscopy (TEM) analyses confirmed that films were composed of a mixture of fine grains of A15 crystalline phase and amorphous material. Elemental chemical analysis also revealed that the films were contaminated with ~ 5at% of oxygen impurities. According to the literature review [29], this appeared to be a dividing impurity concentration suppressing superconductivity. Thus, the later work was focused on reducing oxygen contamination. The source of oxygen background pressure was determined by residual gas analysis measurements and reduced to $6.2*10^{-9}$ Torr by sputtering Ar gas purification and prolonged vacuum chamber plasma cleaning. Despite almost an order of magnitude decrease in oxygen partial pressure, only a minor change in oxygen content was found in the films. Transition temperature measurements did not show superconductivity in films down to 9K. Further reduction of the oxygen content, elimination of the amorphous phase, and increase of the A15 crystalline phase fraction are necessary to demonstrate superconductivity at 4K.

$Nb_3Sn$ films were sputtered on sapphire and Nb substrates and annealed at temperatures ranging from $700^0$C to $1050^0$C. Fabricated samples demonstrated superconducting transitions at temperatures from 7 to 16K. Importantly, one sample of $Nb_3Sn$ coated on Nb substrate and annealed at a relatively low temperature of $700^0$C demonstrated a transition temperature of 14.5K. This suggests feasibility of using LANL sputtering method for producing $Nb_3Sn$ coatings on materials other than Nb, such as copper. The measurements of stoichiometry demonstrated composition of approximately 80 at% of Nb and 20 at% of Sn.

LANL's team also put together a large coating chamber that is capable of producing $Nb_3Ge$ or $Nb_3Sn$ coating inside of an actual 1.3 GHz SRF accelerating cavity. The chamber was set up with a movable target sputter deposition that facilitates achieving uniform conformal coatings on inner surfaces of SRF cavities. Assuming availability of funding, the goal of future work will be the refinement of the coating process, coating of the actual elliptical 1.3 GHz SRF cavity, and the high gradient testing of this cavity.

## 6. THE BRONZE ROUTE

The Japanese National Institute for Materials Science (NIMS) has invented methods to either produce Nb$_3$Sn on bronze substrate or on Oxygen-free Cu. Both processes build upon the A15 superconducting wire technology and also exploit the heat treatment temperature reduction effect of the Cu as ternary element of the Nb-Sn-Cu phase diagram. It is thought that these processes would be suitable to use on SRF cavities fabricated by hydro-forming. Some efforts have started also in the U.S. where Nb films are deposited on a bronze substrate and subsequently annealed to form the Nb$_3$Sn [33]-[35].

### 6.1. Thick Nb$_3$Sn Layers via Bronze Route at NIMS

NIMS methods reproduce the fabrication model of Nb$_3$Sn wires, which includes billet assembly, electron beam welding, hot extrusion, cold-die drawing, intermediate annealing and heat treatment in inert atmosphere. It is well-known that in the bronze process a thick layer of Nb$_3$Sn can be synthesized by a diffusion reaction between Nb and bronze (Cu-Sn) alloy at temperatures below 700 ºC, as opposed to the Sn vapor diffusion process which requires temperatures in excess of 1,000 ºC. This temperature difference affects the grain size of the Nb$_3$Sn phase, in that the grain size through the bronze process (Fig. 7 (b)) is much finer than that through the Sn vapor diffusion process (Fig. 7 (a)). In addition to improving the superconducting properties of the material, a finer grain size may affect the surface morphology as well, with finer Nb$_3$Sn grains helping to keep the Nb$_3$Sn surface in the cavity smoother [9].

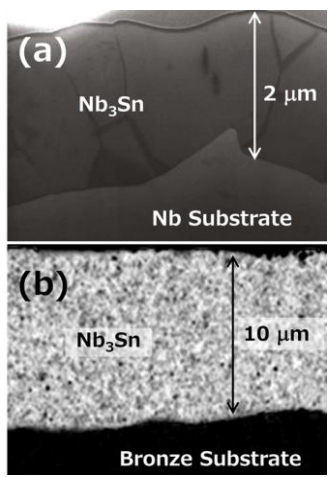
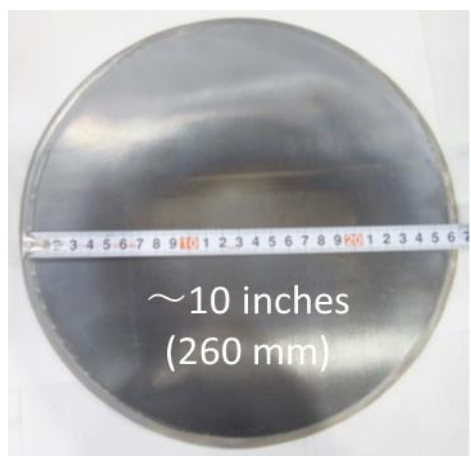

Figure 7: (a) Nb$_3$Sn thin film synthesized by Sn vapor diffusion process [21], (b) Nb$_3$Sn thick layer synthesized by a bronze process.

Figure 8: Nb/Bronze composite plate with a diameter of 260 mm. The thickness of the Nb film is 100 μm.

After some R&D in developing the most successful fabrication method, NIMS has successfully fabricated a Nb/Bronze circular plate 260 mm in diameter and 2.6 mm thick, as shown in Fig. 8. The 100 μm thick Nb film was strongly bonded with the bronze plate by applying a force of 1,200 ton at 720 ºC, using a new precision 1,500 ton hot hydraulic press, as shown in Fig. 9. The next step at NIMS is to form the Nb/Bronze flat-plate into a single cell shape by using a deep deformation press at room temperature as shown in Fig. 10.

The work at NIMS was funded through NIMS Mid-Term Project Research from FY2016-FY2022 under MEXT, and through the TIA Kakehashi projects in FY2021.

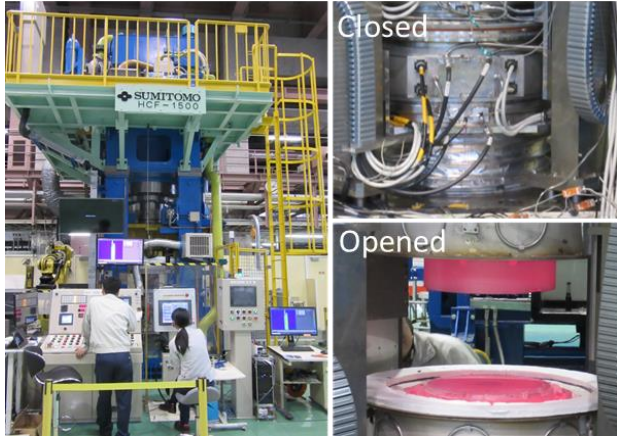
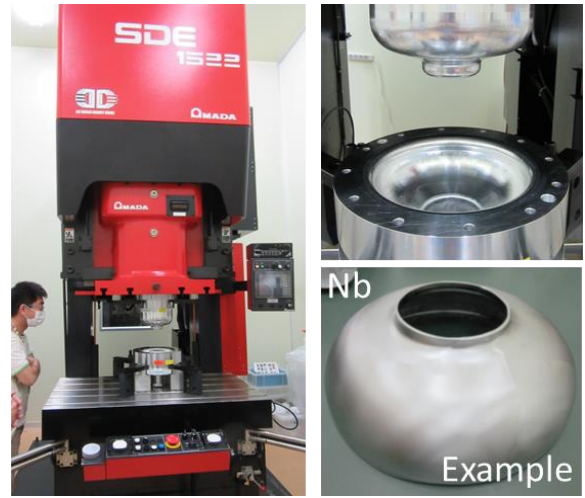

Figure 9: New precision 1,500 ton hot (up to 1,000 ºC) hydraulic press at NIMS.

Figure 10: Deep deformation press to form flat circular plates into single cell shape (KEK owned equipment).

### 6.2. Melt Casting of Bronze Structures in Industry

Energy to Power Solutions (e2P) in collaboration with JLAB have investigated a novel approach to fabricating $Nb_3Sn$ SRF cavities using an ultra-low cost melt casting fabrication process. e2P's simple melt casting techniques can be used to fabricate nearly any superconducting $Nb_3Sn$ structure using either the Bronze Route (BR) or Internal-Tin (IT) processes. The starting bronze substrates were obtained using a two-step fabrication process consisting of first 3D printing a "sand mold" and then subsequently melt casting the bronze substrate using the 3D printed sand mold. Due to the ease of the BR melt casted approach, initial efforts were geared primarily towards this process; however, e2P's patented process [35] can be used for a wide variety of RF structures, including the IT $Nb_3Sn$ fabrication technique as well as normal conducting copper and aluminum conducting cavities.

Multiple 10 mm x 10 mm coupons of varying Sn content ranging from 11 wt% to 19 wt% Sn were heat treated using a modified BR processing technique. All of the coupons were mechanically polished and cleaned prior to Nb film deposition but not chemically etched. The mechanically polished bronze coupons of varying Sn content were initially coated with a 0.5-1 um thick Nb film at JLAB using an Electron Cyclotron Resonance (ECR) RF sputtering technique under varying processing conditions of substrate temperature and incident ion energies. After Nb film deposition, samples were heat treated in a separate furnace to form the superconducting $Nb_3Sn$ phase. Samples were heat treated in both a vacuum furnace at JLAB and inert Ar atmosphere at e2P. Samples heat treated in the vacuum furnace at JLAB showed superior surface smoothness. Post heat treated samples were then tested for $T_{c0}$ using AC susceptibility measurements. Most of the heat treated samples that were examined using SEM/EDX showed a good correlation of $T_{c0}$ onset and transition width ($\Delta T$) of the initial/starting Sn content to the desired stoichiometric $Nb_3Sn$ phase. The lower Sn content coupons resulted in lower $T_{c0}$'s with broader transition widths and the higher Sn coupons resulted in higher $T_{c0}$'s with narrower transition widths. The best samples had $T_{c0}$'s ~17 K and $\Delta T$'s ~ 1 K.

Two samples were further tested for RF surface resistance ($R_s$) and Quality Factor ($Q_0$) at JLAB. These RF measurements were performed at 7.4 GHz using a calorimetric technique and showed a $T_{c0}$ onset ~ 14 K; however, values of $R_s$ were quite disappointing with corresponding $Q_0$ values ~ $10^6$, i.e. nearly two orders of magnitude lower than similar high quality $Nb_3Sn$ films directly deposited substrates by JLAB.

Substantial improvements in the processing variables ranging from higher quality of the initial melt casted structures with higher Sn content, to better polishing and chemical surface treatments will be necessary to realize improved RF performance metrics. Work funded through grant No. DE-SC0018713.

## 7. CONCLUSIONS

High gradient, inexpensive superconducting cavities (SRF) will be needed for future accelerators. However, the accelerating field of Nb cavities is limited by the peak magnetic field on the cavity surface. $Nb_3Sn$ SRF cavities should produce larger accelerating gradients and a larger quality factor, $Q_0$, than Nb cavities. Also, the higher $T_{c0}$ of $Nb_3Sn$ allows cavities to operate at 4.5 K rather than ~2 K that is used for Nb cavities to obtain a higher gradient. This means less expensive refrigeration and more cryogenic reliability.

$Nb_3Sn$ coated SRF cavities, with $Nb_3Sn$ produced by Sn vapor diffusion followed by a thermal reaction at high temperature, have achieved only a fraction of the theoretical predicted gradient. In addition, this process cannot be used on Cu RF cavities. We have presented here instead examples of efforts in developing $Nb_3Sn$ coatings on Cu or bronze. Whereas these methods are all scalable to standard cavity cells, the open-faced Cu RF structures developed at SLAC are the ideal Cu surfaces for coating with superconducting films, as they allow complete access to the inner cavity surface for the coating process, before joining the two half- blocks. Much more progress with respect to that achieved with minimal and often intermittent funding could be made if focus by the funding agencies were put on methods that are promising for coating Cu or bronze instead than on Sn vapor diffusion. If that is done, it is extremely likely that one of these R&D methods will succeed and be available by the same time as the novel concept of parallel-feed RF accelerator structures. A technology that synergistically uses both of these advanced tools would deliver higher gradient and higher temperature of operation and reduce the overall capital and operational costs of any future collider.

*Emanuela Barzi is a Senior Scientist at Fermilab and an Adjunct Professor and Graduate Faculty at OSU. A 2012 Fellow of the APS, a 2021 Fellow of the International Association of Advanced Materials, and a senior member of the IEEE, Barzi has been an active member of the high-energy accelerator and physics communities for 25 years. The Superconducting R&D lab that she founded at FNAL is a world leading research center in low- and high-temperature superconductor technologies for the next generation of particle accelerators. Barzi is a member of the FNAL team that last year produced a world-record field of 14.6 T for a $Nb_3Sn$ accelerator dipole magnet, is co-leading the multi-lab effort on $Nb_3Sn$ undulators for ANL Advanced Photon Source, and is FNAL coordinator of four trilateral EU-US-Japan collaborations. She has co-authored 240+ peer-reviewed papers and book chapters with 6000+ citations. As a member of the Muon g-2 Collaboration, this includes the 2021 paper announcing the difference of the measured magnetic moment from theory expectations. In 2010 she was awarded the Japanese "Superconductor Science and Technology Prize." Barzi also established extensive educational programs at FNAL for graduate students in Physics and Engineering, including the Italian Graduate Students Program, that have benefited hundreds of young professionals, and has mentored 30+ students in her lab for internships, Masters and PhDs. A former councilor of the APS FIP, a former member of the APS Council Steering Committee, and a former Chair of the APS Committee on Prizes and Awards, she is presently a member of the APS Ethics Committee.*